# Investigation of Ancient DNA from Western Siberia and the Sargat Culture

Casey C. Bennett (Department of Anthropology, Indiana University-Bloomington; Centerstone Research Institute[1] ), Frederika A. Kaestle (Department of Anthropology and Institute of Molecular Biology, Indiana University-Bloomington)

Corresponding author:

Casey Bennett

Centerstone Research Institute

1101 6th Avenue North

Nashville, TN 37208

Phone: (615) 460-4111

casey.bennett@centerstoneresearch.org

Keywords: d-loop, Russia, Siberia, aDNA, mtDNA, Sargat, Kurgan

[1] Current Affiliation

**Abstract**

Mitochondrial DNA from fourteen archaeological samples at the Ural State University in Yekaterinburg, Russia was extracted to test the feasibility of ancient DNA work on their collection. These samples come from a number of sites that fall into two groupings. Seven samples are from three sites that belong to a northern group of what are thought to be Ugrians dating to the $8^{th}$-$12^{th}$ century AD, who lived along the Ural Mountains in northwestern Siberia. The remaining seven samples are from two sites that belong to a southern group representing the Sargat culture, dating between roughly the $5^{th}$ century BC and the $5^{th}$ century AD, from southwestern Siberia near the Ural Mountains and the present-day Kazakhstan border. The samples derived from several burial types, including kurgan burials. They also represented a number of different skeletal elements, as well as a range of observed preservation. The northern sites repeatedly failed to amplify after multiple extraction and amplification attempts, but the samples from the southern sites were successfully extracted and amplified. The sequences obtained from the southern sites support the hypothesis that the Sargat culture was a potential zone of intermixture between native Ugrian and/or Siberian populations and steppe peoples from the South, possibly early Iranian or Indo-Iranian, which has been previously suggested by archaeological analysis.

**Introduction**

Human habitation of Western Siberia dates back at least to the late Paleolithic. The region has long-formed an identifiable cultural area marked by similar material culture, including such aspects as tools, weapons, dwellings, and ceramics, which typically are reflective of the environment of the region and hunting and fishing subsistence activities. The area also has a number of major rivers that create a network – the Ob, Irtysh, Ishim, and Tobol for instance – providing transportation routes and opportunities for social, cultural, and economic links. By the Late Neolithic, fortified settlements began to appear, with increased social and economic complexity evidenced by higher-density settlements, indications of warfare, and intensive use of local raw materials and expanding trade networks. With the onset of the Eneolithic, Bronze Age, and Iron Age, the Ural Mountains and the surrounding areas became important centers for metallurgical resources. This, along with the expansion of emerging steppe cultures, eventually led to increasing interaction of the regions around the Urals with their neighbors to the south on the steppes of Central Eurasia (Koryakova and Daire 2000, Andrey Shpitonkov et al., Ural State University, Personal Communication 2004). This development is evidenced as early as 3000 BC in the Botai and Surtanda cultures of the southern Urals and northern Kazakhstan. These cultures have evidence of metal mining (Matyushin 1986) and horseback riding, though they seem to lack wheeled vehicles or formal cemeteries (Anthony 1995, 1998). The eventual onset of the Iron Age in the area marked a disintegration of some earlier cultural units and a period of shifting cultural traits, cultural borrowing, and cultural admixture (Matveeva 2000, Andrey Shpitonkov et al., Ural State University, Personal Communication 2004). It is also from the Trans-Ural area that modern Ugrian populations are thought to have derived (Golden 1991, Andrey Shpitonkov et al., Ural State University, Personal Communication 2004).

On the steppes around this time during the 3rd-1st Millennia BC, loosely defined archaeological cultures begin emerging, such as the Sintashta-Petrovka (2100-1700 BC) and the Andronovo and Srubnaya of the 2nd millennium BC, and the later Saka populations of the Kazakh steppes and surrounding areas in the 1st millennium BC (as well as other groups such as the Sarmatians and Scythians to the West and the Karsuk to the East). These groups are considered to be Iranian or Indo-Iranian, and at least partly nomadic pastoralists (though their economy was likely much more complex and variable than this generalization) (Mallory 1989, Sinor 1991, Anthony and Vinogradov 1995, Mair 1998).

The Sargat culture was located in the forest-steppe region of Southwestern Siberia, near what is now the border of Russia and northern Kazakhstan, from around the 5th century BC till the 5th century AD. It is associated with a number of similar archaeological cultures in the region from the same period or slightly preceding it, for example the Gorokhovo, Iktul, and Baitovo. It is also known for containing a number of Kurgan burials (Koryakova and Daire 2000, Matveeva 2000, Andrey Shpitonkov et al., Ural State University, Personal Communication 2004), and roughly half of all graves contain the remains of horse harnesses (Koryakova 2000). The Sargat Culture, based on archaeological evidence, has been ascribed to a zone of intermixture between the Iranian steppe peoples to the South, such as the Saka or Sarmatians and the native Ugrian and/or Siberian populations (Koryakova and Daire 2000, Matveeva 2000, Andrey Shpitonkov et al., Ural State University, Personal Communication 2004). Previous craniological research has also suggested some intrusion of Iranian peoples from the South (Matveeva 2000).

The collection of the Ural State University in Yekaterinburg, Russia contains archaeological samples from sites in the region representing a range of these historical periods.

A number of the samples were examined to test the feasibility of ancient DNA analysis. By design, the sample set included a variety of skeletal elements (including teeth, ribs, and vertebrae, as well as a metatarsal and talus) with varying degrees of preservation. These samples came from several sites that fell into two groupings: a northern group and a southern group. The northern group is thought to represent Ugrians and dates to the 8th-12th centuries AD from along the Urals in northwestern Siberia. This group included the sites of Sagyatino 6 (two samples, near Surgut), Endyrsky II (three samples, lying between the Lower Ob and the Urals), and Zeleniy Yar (two samples, near Salekhard and the mouth of the Ob). These are represented on the map in Figure One.

The southern group represents the Sargat culture dating between roughly the 5th century BC and the 5th century AD from southwestern Siberia near the Urals and the Kazakhstan border. There were three samples from Kurtuguz I, deriving from a multi-skeleton pit burial. There were four samples from Sopininsky, representing two individual graves, one being a kurgan burial mound and the other being a "flat" or "ground" grave that was discovered between the kurgan mounds. These two sites are represented on the map (Fig.1) as the more southerly circle south and east of Yekaterinburg.

**Materials and Methods**

The laboratory work was performed at the Indiana Molecular Biology Institute (IMBI) at Indiana University, Bloomington in the Kaestle Lab. The skeletal samples were UV irradiated (254 nm at a distance of 5 inches) for 5 minutes each side, soaked in 6% sodium hypochlorite for 10-15 minutes, and rinsed in molecular-grade water (Eppendorf) for 10-15 minutes. Half a gram of each sample was decalcified in 2 ml of 0.5 M EDTA, pH 8.0, with rocking across three days at 40° C. After centrifugation, the EDTA was replaced, 200 μL of 20 mg/ml Proteinase K was added, and the samples were digested overnight at 50° C with rocking. The samples were then processed using the GeneClean ancient DNA Kit (MP Biomedical) to extract the DNA, using the manufacturer's protocol with the elimination of the Dehybernation Solution step. The samples were then filtered and concentrated using Centricon-100 units (Millipore). All extractions included a negative control.

The extracts were amplified in 25 μL reactions, using DNA-free ddH$_2$O (Eppendorf), 0.2 mM of each dNTP, 1x PCR buffer (Eppendorf), 1 mg/mL BSA, 1.5 Units Hotmaster Taq (Eppendorf), 0.24 μM of each primer, and 3 μL of ancient DNA extract. These were then denatured at 94° C for 10 min, followed by 43 cycles of 94° C for 30 sec, 55° C for 30 sec, 72° C for 30 sec, and a final extension at 72° C for 5 min. All amplifications included a negative control.

The first hypervariable segment (HVSI) of the mitochondrial D loop was amplified in three overlapping fragments using the primer pairs L16038/H16213, L16209/H16356, and L16347/H16454. The final segment was also amplified using H16508 rather than H16454 in an attempt to alleviate amplification/sequencing problems in the third segment.

Amplification products were visualized with ethidium bromide on 2% agarose gels and purified using the Qiaquick PCR Purification Kit according to manufacturer's protocols (Qiagen). Sequencing reactions were performed using the ABI Prism BigDye 3.1 kit (Applied Biosystems) according to manufacturer conditions and analyzed on an ABI 3730 Sequence Analysis System (Applied Biosystems). Sequences were corrected and compared against the Revised Cambridge Reference Sequence (Anderson 1981; Andrews et al., 1999) using Sequencher 4.1 (Gene Codes Corp). All fragments were sequenced in both the forward and reverse direction from at least two separate extractions.

The extractions and amplifications were performed in a positive pressure, physically isolated, ancient DNA-dedicated laboratory, which is UV irradiated and bleached regularly. All post-PCR analysis is performed in a separate lab. All lab personnel wear disposable lab coats, face masks, gloves, and sleeve guards, and travel between the ancient and post-PCR laboratories is unidirectional in any single day. In addition, the sequences of all lab and maintenance personnel have been identified. Extraction and amplification negative controls were performed in all cases.

All final consensus sequences were aligned and compiled using Sequencher 4.1 (Gene Codes Corp). The final sequences were then compared to an aggregate list of identified haplogroups compiled from the literature (Sykes et al. 1995, Macauley et al. 1999, Redd and Stoneking 1999, Helgason et al. 2000, Richards et al. 2000, Simoni et al. 2000 and references therein). Further comparisons were performed using unpublished data kindly provided by Toomas Kivisild and Mait Metspalu (Metspalu et al. 2004), and supplemented with additional data from GenBank (Benson et al. 2000), as detailed in Bennett and Kaestle (2006).

**Results**

Attempts to extract and amplify DNA from the northern site samples repeatedly failed, although between four and eleven amplification attempts were made per sample (for a combined total of 49 amplification attempts from northern site samples). It should be noted that the reactions of several of the northern samples to the chemicals used during extraction was atypical. This included changes in color (to a bright blue-green) and malleability in some of the samples from Sagyatino and Zeleniy Yar. None of these reactions was seen in any of the southern samples. One of the possible explanations for some of these reactions may be exposure to metal from the surrounding soil or perhaps some sort of mineralization. Due to the unusual reactions to the extraction reagents, some of the northern site samples were only extracted once.

The northern sites included a number of different skeletal elements, indicating that the failure was not related to the choice of sample. This included a rib, tooth, and tallus from Endyrsky (graves 12, 15, and 19 respectively), two vertebrae from Sagyatino (graves 82 and 89), and a tooth and vertebra from Zeleniy Yar (graves 1 and 34). Although some samples were extracted only once (see above), their failure to amplify cannot be attributed to a failed extraction because other ancient samples extracted at the same time were successfully analyzed. It should also be noted that one of the samples from Zeleniy Yar, ZJ1, included some desiccated soft tissue remaining on the skeletal elements, which were removed prior to extraction. Thus, the failure to amplify DNA from these samples was not due to overall gross preservation. In order to determine if the DNA extracts from these northern samples contained co-extracted PCR-inhibiting compounds, these extracts were added to HVSI amplifications of modern mitochondrial DNA samples, but no inhibition was detected (data not shown).

The southern sites were both successful in all phases to varying degrees. The results can be seen in table 1. The three Kurtuguz individuals belonged to haplogroups A, C, and Z. Two of the samples (K3T and K3R) were originally thought to derive from the same skeleton but upon final analysis were determined to be from different individuals. This result was replicated and verified. All three samples derived from a single multi-skeleton pit burial at the same site, and it is possible that skeletal remains were not originally separated or were later disturbed (as K3T is a tooth sample and K3R is a rib sample). All three samples gave clean, uncontaminated distinct sequences, indicating that they represent three distinct individuals.

The four Sopinisky samples represented two different graves, corresponding to two individuals. The kurgan burial included a tooth and rib sample, which resulted in a sequence belonging to haplogroup T (more specifically T1). This sequence was a relatively uncommon variant of T/T1 having the mutation 16243C. The flat grave included a tooth and metatarsal sample, which both yielded sequence belonging to haplogroup Z. The haplotype of the individual in the flat grave, however, was different from the haplotype of K3R (which was also haplogroup Z).

In all of these samples, sequencing of the third segment (16347-16454) was ultimately unsuccessful. Though some results were obtained for a few of the samples, there were problems verifying the authenticity (due to the lack of any differences from the Cambridge Reference Sequence in the segment for any of the obtained sequences). Different primers were used to alleviate the problem without any success. Given the difficulty of obtaining replicable results for samples from these sites in this particular segment, as well as issues with sequence verification, sequence from this segment was not included in the final analyses. Furthermore, amplification of the 16038-16213 fragment was unsuccessful for the rib sample K2 from Kurtuguz despite repeated attempts. This sample was generally more difficult to amplify than the other southern samples (with only a 36% success rate, compared to an overall southern amplification success rate of 70%). Thus all sequences obtained represent the segment 16038-16356 (319 base pairs), except for K2 (16209-16356, 148 bp).

Overall, the most successful samples were the teeth, which consistently produced strong amplifications and clean sequence, while there were some difficulties in working with the rib and metatarsal samples. Among the southern samples, the success rate for teeth was 32/40 amplification attempts (80%), while the success rate for bone was 31/50 amplification attempts (62%). However, positive results were obtained from all southern samples given sufficient repeated attempts. It should also be noted that outward signs of preservation (as judged by the authors) were not indicative in any way of the ultimate success of the samples in genetic analysis, consistent with the results from the northern samples.

**Discussion**

It is difficult given the limitations of the haplogroup approach to infer specific population history. As such, this study represents only the first step in reconstructing population affinities. T is a widespread haplogroup throughout Western and Central Eurasia with varying degrees of prevalence and certainly might have been present in other groups from the surrounding areas or in the Sargat territory itself. The T haplogroup is found in roughly 8% of modern Europeans (Bermisheva et al. 2003). It is also common among modern day Iranians. Based on a sample of over 400 modern day Iranians (Kivisild and Metspalu, Estonian Biocenter, personal communication 2003), the T haplogroup represents roughly 8.3% of the population (about 1 out of 12 individuals), with the more specific T1 subtype constituting roughly half of those (see Figure 2). On the other hand, T is thought to have emanated from the Middle East (Bermisheva et al. 2002). Furthermore, the specific subtype T1 tends to be found further east and is common in Central Asian and modern Turkic populations (Lalueza-Fox et al. 2004), who inhabit much of the same territory as the ancient Saka, Sarmatian, Andronovo, and other putative Iranian peoples of the 2$^{nd}$ and 1$^{st}$ millennia BC. Lalueza-Fox et al. (2004) also found several T and T1 sequences in ancient burials, including kurgans, in the Kazakh steppe between the 14$^{th}$-10$^{th}$ centuries BC, as well as later into the 1$^{st}$ Millenia BC. These coincide with the latter part of the Andronovo period and the Saka period in the region.

Further evidence pointing toward genetic influence from the South comes from analyses of populations to the west. Although haplogroup T exists in modern Russian and Belorussian populations at a frequency of roughly 7% (Belyaeva et al. 2003), the Slavic expansions east and north, particularly as far east as the trans-Ural region, probably came after or in the latter stages of the Sargat period (Mallory 1989, Moss 2001, Malyachurk et al. 2005). Beyond that, the exact composition and genetic structure of populations that may have preceded them are unclear, though they were probably of a Finno-Ugrian nature (Moss 2001).  Analysis of Saami populations in Scandinavia found virtually no haplotype T despite the presence of other European haplotypes, though the Saami may be a special case given the presence of T in modern Finns (Tambets et al. 2004).  A more general comparison by Bermisheva (2002) found that, at least maternally, "the Finno-Ugric populations of the (Ural) region proved to be more similar to their Turkic neighbors than to linguistically related Balto-Finnish ethnic groups" (p. 82).  The genetic findings in that study pointed toward a joint role of Siberian and Central Asian influence in the ethnogenesis of modern trans-Ural populations (Bermisheva 2002).

Further analysis from Sargat sites will likely clarify this issue.  Nevertheless, given the existing archaeological evidence in a number of sources (Koryakova and Daire 2000, Matveeva 2000, Andrey Shpitonkov et al., Ural State University, personal communication 2004), the known mitochondrial haplogroup prevalence among modern day populations, and the distribution of kurgan burial traditions in Central Eurasia, it is reasonable to assume for the time being that the individual from the kurgan burial at Sopinisky was of Iranian descent from the steppes to the south, possibly from the Saka or another Iranian group, such as the Sarmatians.  Whether this is related to the underlying genetic structure of modern populations in the trans-Ural region remains to be seen.

The haplogroups of the other samples, A, C, and the two variants of Z, are typical of Siberian populations.  A, C, and Z are common in Northern Asia, particularly north of the Altai Mountains and the Amur River (i.e. Siberia), and they decrease in frequency as you move south, with Z being rare at best (as one might expect, there is 1 individual of haplogroup Z present in the Iranian sample discussed above, 3 members of haplogroup C, and only three with a variant of haplogroup A). In fact, A, C, and Z along with D, G, and Y constitute approximately 75% of the haplogroups of Siberia (Mishmar et al. 2002, Derenko et al. 2007). Haplogroup Z is also found farther west across the northern tier of Eurasia into Scandinavia. Z is also present in modern day Ugrian populations, including those in the Ural region, ranging from 0-5% of sampled populations. Haplogroups A and C also exist in these and Turkic populations to varying degrees (Comas et al. 1998, Bermisheva et al. 2003). However, an analysis of ancient Kazakhs found that East Asian haplogroups such as A and C did not begin to move into the Kazakh steppe region till around the time of the Hsiung-Nu ($1^{st}$ Millenia BC), which is around the onset of the Sargat Culture as well (Lalueza-Fox et al. 2004).

The presence of these haplogroups in burials at sites within the Sargat culture, and moreover in a flat grave at the same site with a kurgan burial of haplogroup T, is consistent with the hypothesis that the Sargat culture was a zone of intermixture, as suggested by archaeological evidence. While it is not discernible if the native population present in the Sargat population is specifically Ugrian, evidence of Iranian loan words in old Ugrian is suggestive of some interaction (Golden 1991), whether it happened in the Sargat culture or elsewhere, or indirectly via other Siberian populations, such as those who may have been present in the area of the Sargat culture at the time. This issue might be clarified if successful DNA extraction from the northern sites could be achieved. Either way, it is clear that cultural admixture was ongoing in the Sargat culture, and the preliminary genetic evidence appears to support the hypothesis that that process was reflective of movements of biological populations in the region at the time.

The haplotypes of the samples from Kurtuguz and Sopinisky seem consistent with earlier suggestions based upon the archaeological evidence that the Sargat culture was a zone of intermixture between the native Ugrian and/or Siberian populations and expanding steppe cultures to the south. Matveeva (2000) suggested this development was via an *elite dominance* process, with different components of the Sargat population having different origins and varying social rank. As the kurgan burial was not only a cultural trait common in the Central Eurasian steppes but also a symbol of status, the results from this study might support such an argument. The individual belonging to haplogroup T may represent the northward movement of Iranian steppe peoples and perhaps some of the sociopolitical realities at the time.

**Acknowledgments**

The authors would like to thank the faculty and staff of the Ural State University and the Ugrian Research Center in Yekaterinburg, including Andrey Shpitonkov, Dr. Ludmila Koryakova, Dr. Dima Rajev, and associates, for making available their extensive collection and aiding us in this research.  The authors would also like acknowledge Dr. Bryan Hanks of the University of Pittsburgh for his special assistance in facilitating this study, and Christina Van Regenmorter at the Centerstone Research Institute for draft review assistance.  Special thanks to Dr. Toomas Kivisild and Mait Metspalu of the Estonian Biocenter for providing their data for our use. This research was funded by and performed at Indiana University, Bloomington in collaboration with the Indiana Molecular Biology Institute (IMBI).

Figure 1: Distribution of Sample Sites )

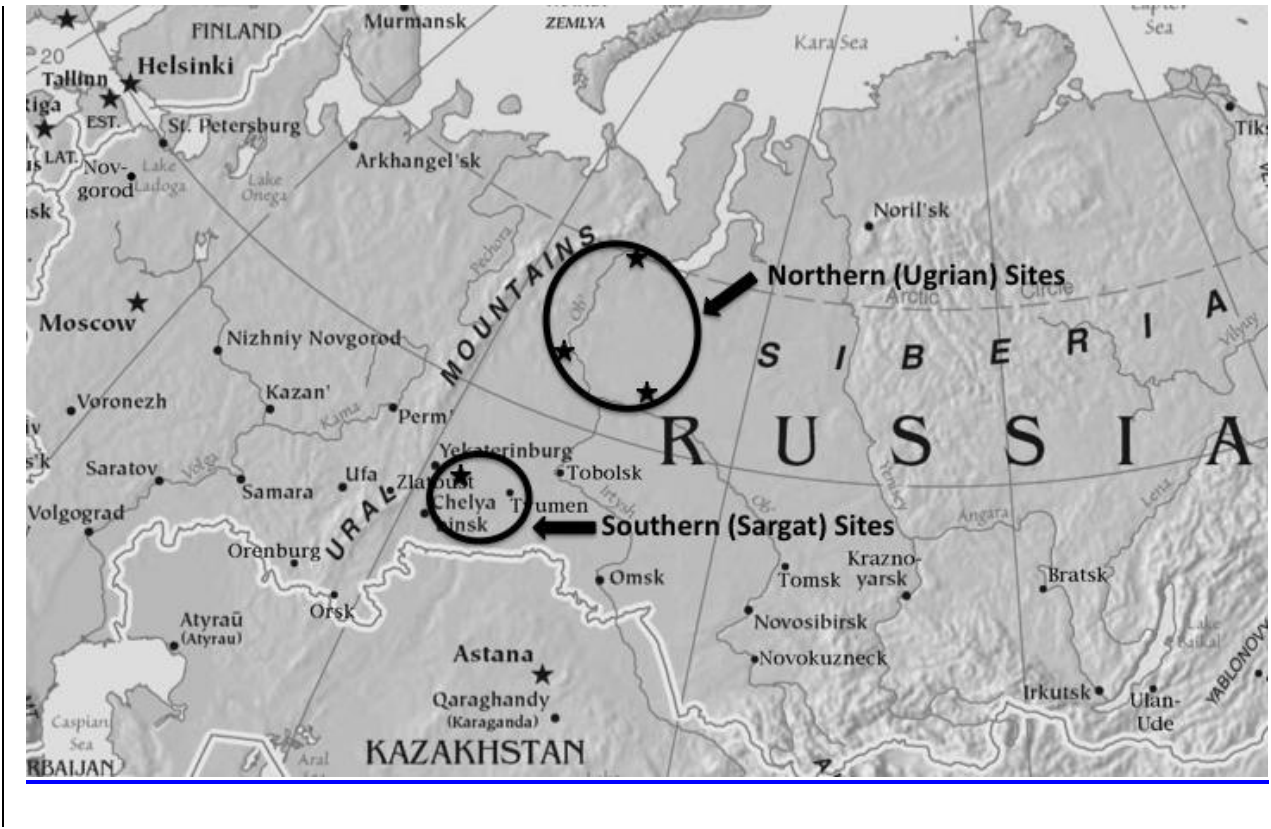

Table 1: Mutation Table for Sargat Samples, with the Revised Cambridge Reference sequence (Anderson et al. 1981; Andrews et al 1999) listed for comparison. Nucleotide positions minus 16000 are listed in the top row, and haplogroup assignment in the final column.

|  | 93 | 126 | 129 | 163 | 185 | 186 | 189 | 215 | 223 | 224 | 243 | 260 | 290 | 294 | 298 | 319 | 327 | 335 | Hap |
|---|---|---|---|---|---|---|---|---|---|---|---|---|---|---|---|---|---|---|---|
| Reference | T | T | G | A | C | C | T | A | C | T | T | C | C | C | T | G | C | A | H |
| **Sopinisky** | | | | | | | | | | | | | | | | | | | |
| SK[a] | * | C | * | G | * | T | C | * | * | * | C | * | * | T | * | * | * | * | T1 |
| SF[b] | * | * | A | * | T | * | * | G | T | Y | * | T | * | * | C | * | * | * | Z |
|  | | | | | | | | | | | | | | | | | | | |
| **Kurtuguz** | | | | | | | | | | | | | | | | | | | |
| K2 | ? | ? | ? | ? | ? | ? | ? | * | T | * | * | * | T | * | * | A | * | G | A |
| K3T | * | * | * | * | * | * | * | * | T | * | * | * | * | * | C | * | T | * | C |
| K3R | C | * | * | * | T | * | * | * | T | * | * | T | * | * | C | * | * | * | Z |

[a] Kurgan Burial

[b] Flat Grave Burial

Figure 2: Eurasian Haplogroup Distribution

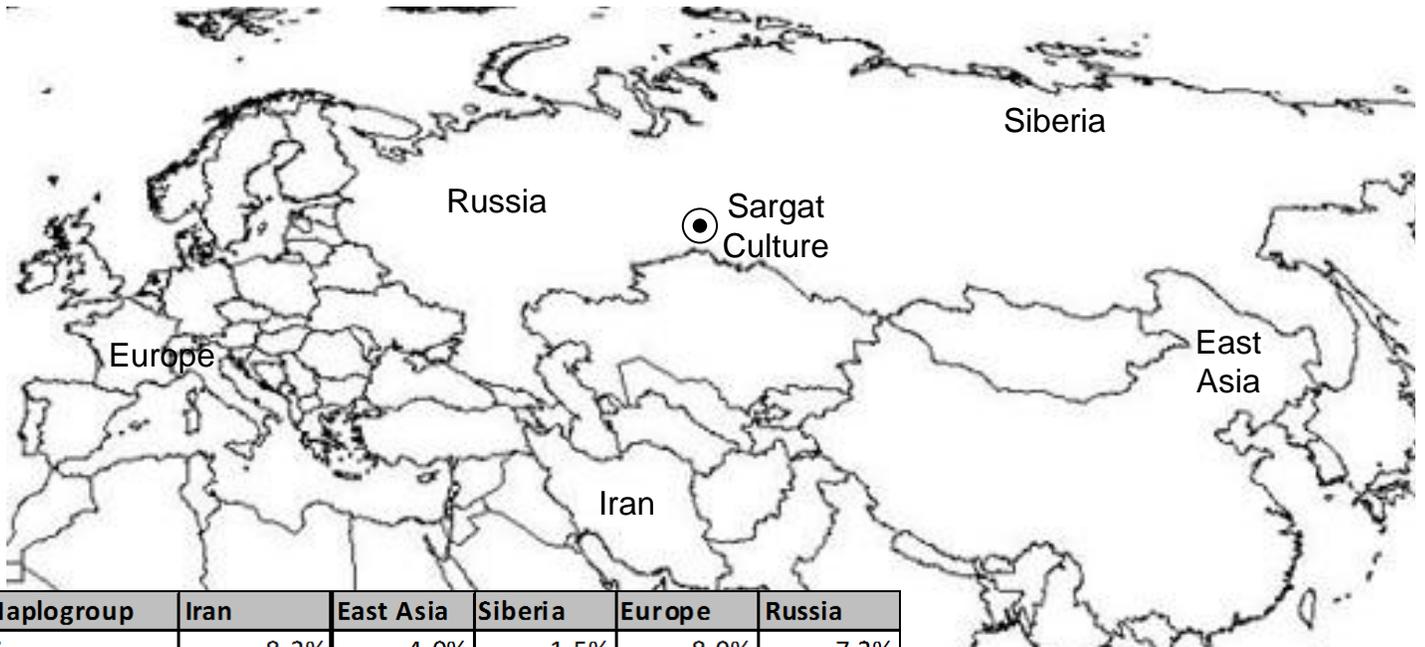

| Haplogroup | Iran | East Asia | Siberia | Europe | Russia |
|---|---|---|---|---|---|
| T | 8.3% | 4.0% | 1.5% | 8.0% | 7.2% |
| A, C, Z | 1.8% | 16.4% | 41.6% | 0.5% | 0.8% |
| Other | 89.9% | 79.6% | 56.9% | 91.5% | 92.0% |